# A Pandemic-Resilient Open-Inquiry Physical Science Lab Course Which Leverages the Maker Movement

F. R. Bradbury, Amsterdam University College, Amsterdam, The Netherlands, F.R.Bradbury@auc.nl

C. F. J. Pols, University of Technology Delft, Delft, The Netherlands, C.F.J.Pols@tudelft.nl

**Abstract:**

Without any major changes, a pilot version of a physical science lab course was able to continue when the COVID-19 crisis necessitated the abrupt suspension of on-campus education.  The 'Maker Lab' course, in which students conceive and set up their own experiments using affordable microcontrollers, required students to follow the entire arc of an empirical research cycle twice.  Pedagogical literature on teaching the process of experimental research and the scientific method motivate use of these *open-inquiry* assignments. Further, the *flipped classroom approach* was used, where contact time is devoted to discussions and the students' actual experiments were carried out independently at home or elsewhere without the supervision of an instructor.  Despite the COVID-19 measures, all students were able to produce interesting and successful research projects.  While there were of course difficulties encountered due to the abrupt transition to online teaching, we found several counterbalancing advantages that bear consideration for inclusion even when all teaching activities can return to campus.  We believe that three components in the design of the course were vital to the resilience of the course:  the choice for fully open-inquiry projects, the decision to use Arduinos as measurement tools, and the flipped aspect of the instruction methods.  We also include considerations for adapting these pandemic-resilient methods in other courses and programs.

# Introduction

The ongoing COVID-19 pandemic and the associated social distancing measures have required many universities to quickly adjust to distance education methods, an adjustment which is especially difficult for science lab courses.  We report on a new physical science lab course piloted this spring at Amsterdam University College (AUC) which was fortuitously found to be pandemic-resilient.  The 'Maker Lab' course aims to present students with a different way of learning skills in experimental science via open-inquiry projects that follow the entire arc of an empirical research cycle using widely available microcontrollers.  It is cross-listed between AUC's physics and information science tracks, and is open to all natural science students who have completed their first year mathematics requirements.

Piloting a totally new course on scientific inquiry at university level for students of different backgrounds is already ambitious.  But on March 13$^{th}$ the instructors and students encountered an extra challenge:  the university buildings closed for the remainder of the semester.  It is purely coincidence that this pilot was running at the time of the COVID-19 outbreak, but with hindsight, the course design played a critical role in the course's pandemic-resilience.

Although the course's methods and results must still be rigorously evaluated, we can present our preliminary observations that point toward the validation of our methods for achieving the intended learning outcomes.  This article focuses on the course's pandemic resilience, and specifically which features allowed it to continue without any significant changes.  Our methods can help other practitioners in switching to pandemic-resilient open-inquiry approaches in a broad selection of science lab courses.

# Problem situation

In a traditional science lab course, students do many of the easiest tasks of a particular experimental project (e.g. setup the materials and take repeated observations) while under instructor supervision in a laboratory.  Traditional "confirmation" and "structured inquiry" activities (Table 1) do not ask the students to practice defining a feasible research question nor design a procedure for investigating it.  Students are typically sent home to do the difficult work of analyzing and drawing conclusions from their data.  It is rarely seen that students are asked to evaluate the reliability of their work (Hodson, 1990; Holmes & Wieman, 2016).

Table 1:  The levels of inquiry in lab assignments are determined by how much of an experiment's components are given or specified by the instructor (Tamir, 1991).

|  | **Problem** | **Procedure** | **Conclusion** |
| --- | --- | --- | --- |
| **Confirmation** | given | given | given |
| **Structured inquiry** | given | given | open |
| **Guided inquiry** | given | open | open |
| **Open inquiry** | open | open | open |

Students from AUC, to a large extent, devise their own curriculum by choosing their own courses. A restriction is that natural science students must include at least one dedicated lab course in their curriculum. As students do not engage in many lab courses, a main goal of our four month, six ECTS credit course is that students with very different educational backgrounds should authentically experience the empirical research process firsthand. This requires moving away from traditional "cookbook" experiments, where a step-by-step recipe is given which students must complete. This aim is in line with recent pedagogical research (Ansell & Selen, 2016; Wilcox & Lewandowski, 2016; Zwickl, Hirokawa, Finkelstein, & Lewandowski, 2014) showing that moving away from traditional "cookbook" experiments leads to more effective teaching of the process of experimental research and the methods of scientific research, while not disadvantaging students' conceptual learning outcomes (Holmes, Olsen, Thomas, & Wieman, 2017; Wilcox & Lewandowski, 2017). These shifts involve requiring students to design their experimental methods ("guided inquiry") or even to choose their topic and their research question (fully "open inquiry"), thus achieving a more authentic experience in conducting experimental research, see Table 1.

A constraint at AUC is the college's lack of in-house laboratory facilities. Lab courses typically consume a substantial amount of a department's educational budget as special instruments and lab rooms are required. In this case, our liberal arts and sciences honors college pays other departments for using their facilities for our lab courses. Before the Maker Lab pilot, AUC had provided only one lab course per natural science discipline.

The course is thus designed with the aim of teaching scientific inquiry in a flipped classroom style using affordable and accessible materials.

## Course design

To support students in navigating the full research cycle, the *most open variant of inquiry* in Table 1 was chosen. The choice is further warranted by pedagogical research showing that open-inquiry methods outperform guided-inquiry in teaching students the strategies, attitudes, and habits of mind which are advantageous in experimental science (Wilcox & Lewandowski, 2016), well as research showing that open-inquiry activities address the full variety of cognitive tasks necessary for an empirical research cycle (Holmes & Wieman, 2016; Wieman, 2015). In order to give students sufficient time for design, reflection, and iteration in their open-inquiry projects, we heeded literature suggesting that more time devoted to fewer projects allows for greater learning outcomes (Luckie et al., 2012).

Open inquiry also has consequences for student ownership. When given sufficient support to carry it through, student choice of their research topic usually increases student motivation (Hodson, 2014). We give students opportunity and support in defining their project topic, a specific research question, and the methods for measurement and data analysis. The inquiry is their own, and success and failure are framed in terms of their own goals. With the responsibility in their hands, their critical engagement is ensured at every step in the research cycle. Finally, the students' freedom in choosing project topics allows the Maker Lab to indeed cater to an interdisciplinary group of natural science majors.

Partly to address the lack of facilities and partly to enable fully open inquiries, we applied the idea of "*flipped classrooms*" in our course design, whereby the students' actual measurements are conducted outside of a supervised classroom environment. This allows the instructor contact time to focus on the most challenging activities (Hodson, 2014) (e.g. experimental design, data and uncertainty analysis, iterative improvements for advancing scientific evidence). Instructor-student discussions focus on goal-setting, planning, and evaluating findings; and the students hold chief responsibility in all of these discussions for effectively communicating their ideas and work.

When choosing for open-inquiry and flipped classroom approaches, there are associated constraints on students' inquiries regarding the choice of their instruments. The experimental tools must be cheap enough to accept the risk of being damaged, sufficiently flexible and accessible to be of value in the hands of a diverse and inexperienced group of students, and inherently safe for un-supervised use. To more fully empower students in the experimental design process, these tools must allow for a sufficient level of control and insight into their workings. A pre-calibrated "black-box" measurement device affords little or no control of - and often scant information on - its readout uncertainties and their possible effects on students' measurements. Rather, building their experimental tools from more basic components gives students fuller control and insight into measurement calibration and uncertainties (Bouquet, Bobroff, Fuchs-Gallezot, & Maurines, 2017). As a consequence, students better understand the accuracy and limitations of their data and their validity in being used as scientific evidence.

We chose Arduinos and the myriad variety of sensors they can control because they satisfy all of the above-mentioned requirements. Students receive a short training in using Arduinos to control and read electronic sensors and are further directed to consult user-friendly sensor documentation, Maker forums, and online instructional videos; all of which are easily discoverable and accessible thanks to the open-source ethos of the mature and dynamic Maker movement.

## Results

After receiving substantial training on Maker skills & data analysis via short structured experiments which are designed to explicitly motivate steps in the empirical research process, students dedicated much time (two thirds of the course) to two successive open-inquiry projects, performed in pairs. In the first round, students posed questions addressing:

- building and improving the signal processing of an Arduino theremin
- comparing water retention of alternative potting soils against those with unsustainably harvested peat-moss
- optically measuring heart rate and characterizing its post-exercise recovery to equilibrium
- measuring color fidelity of a Macbook's screen with an RGB sensor
- investigating a photovoltaic cell's power's dependence on illumination angle
- pushing the Arduino's sampling rate for precision sound frequency determination
- measuring local wind-speeds to determine suitable bee-habitat

The COVID-19 crisis started soon after the start of the first round projects, but the students already had their most important tools in hand, thus their projects could continue without any structural problems. It should be noted, however, that this first round occurred during the most stressful period of the pandemic, and for this reason several students were delayed in submitting their first round assignments.  The second round of open-inquiry projects were completed start-to-finish under the pandemic's social distancing restrictions.

The only adaptations in project scope involved just one student due to the severe lock-down restrictions they faced back home in South Africa.  It was decided to allow this student to use sensors in a smartphone instead of controlled directly by Arduinos because of the limited material that they had been able to take home, and also to work individually instead of in a pair.

The inquiry process for the second round projects was observed to be generally much smoother for students than the first round, where there were a couple false-starts and some misunderstandings in the construction of scientific evidence.  Judging by the improvements in the second round, the feedback provided on the first projects and/or the experience the students had gained must have been effective.  The second round projects addressed the following topics:

- comparing signatures of bicep muscle fatigue between dominant and non-dominant arms with median frequency evolution of the electromyography (EMG) power spectrum
- building and characterizing performance of a swiveling Arduino sonar radar
- comparing accelerometer measurements of a beam's fundamental oscillation frequency with the Euler-Bernoulli model
- comparing air pollution levels inside apartments on the road-side and courtyard-side of the student residence building
- investigating whether self-reported joke funniness correlates with EMG signals of facial muscles
- comparing two measurements of bread-dough rise / yeast activity:  $CO_2$ gas sensing & volume changes via ultrasonic ranging
- studying effects of temperature on germination of cress seeds

Following the original plan, most instructor contact hours during the open-inquiry projects were devoted to individual team meetings.  Interestingly, the flipped lab methods and distance learning constraints guaranteed that student projects retained independence.  This physical separation alleviates a concern that our instincts as teachers to help can go too far and counter-productively rob students of ownership, leaving them with an exercise in confirmation or structured inquiry (Hodson, 1990).

We also observed a desired consequence of student ownership related to our open-inquiry and flipped-lab methods:  student communication skills are continuously sharpened as they are wholly responsible for conveying their ideas, reasoning, and observations in their individual meetings with instructors.

To give one concrete example of how students were supported in their open-inquiry process, we look at the top of the list of second round projects:  comparing electrical signatures of muscle fatigue.

> One pair investigated muscle fatigue and how it might be different for different people.  They used electromyography sensor boards that amplify the small voltage differences which are passively picked up by electrodes on the skin.  They reviewed literature and found that a muscle's EMG signal's median frequency (defined via its power spectrum) typically decreases as it becomes fatigued.  They proposed a simple static hold experiment which eliminated any complications involved in muscle movement.  Upon explaining their ideas, they were given the go-ahead and they ordered the sensor and supplies, and started testing. Before the midway point of the project, in an individual team meeting video call, they demonstrated a live plot of their own EMG data acquired from the bicep.

In subsequent team meeting video calls, they reported on results and were helped with data analysis and drawing conclusions. They mainly needed help to understand and implement the use of a Fourier transform and a power spectrum for finding a median frequency. After seeing a single example, they were able to chop their 18 time series data sets into segments, calculate median frequencies, plot their evolutions during the muscle fatigue experiments, and check in briefly about whether their observations and conclusions sounded reasonable.

Similar to this example, all of the second round projects exhibited substantial student independence and rational progression through the empirical research process.

While full course evaluation will not be complete until later this summer, we can already claim success on the most critical measures: the students have succeeded in conceiving, designing, and carrying out a wide variety of experiments suiting their interdisciplinary interests. With the experience of running this pilot, we find that the required initial investment of materials is less than €100 per student and that annual costs thereafter are much lower.

**Factors contributing to pandemic-resiliency**

The Maker Lab's resiliency for continuing with little adjustment despite the transition to distance learning seems exceptional for a science lab course as most are currently being altered in significant ways, delayed, or even canceled. For descriptions of changes to physics labs, see for example (Fox, Werth, Hoehn, & Lewandowski, 2020; Pols, 2020). The only significant changes in instructional and assessment methods were switches to conducting individual team meetings over video call and the replacement of most synchronous in-class presentations with pre-recorded videos, and neither of these small changes seemed to compromise the main learning objectives. In our understanding we see three main factors that contributed to pandemic resiliency: the choice for fully open inquiries, the decision to use Arduinos as measurement tools, and the flipped aspect of the instruction methods. Besides these principal course design choices, it was also advantageous that the course was given in a four month extensive format and that it included two full open-inquiry projects. We briefly discuss these five factors and their contribution to the resiliency.

The most important factor in pandemic-resiliency is that Maker Lab students were tasked with conducting measurements outside of supervised classroom environments. Our subsequent choice to leverage Arduinos and the sensors they can control is important for enabling this flipped lab approach. Thanks in part to their low cost and inherent safety all students could be provided with an Arduino and sensors to conduct their experiments at home, where the accessibility of online Arduino resources supported students in overcoming hurdles and making progress independently. When required, additional sensors and supplies could be ordered and quickly shipped to students' homes directly.

As mentioned above, the only change in instruction was the moving of individual team meetings with the instructor to video calls; and even this change was not actually so negative. As instructors, we found it straightforward and worthwhile to discuss and support students' *own* plans, feasibility estimates, and findings. Supporting open inquiries was thus found to be easier than attempting over video call to guarantee that students were correctly taking all the pre-defined steps and reaching the end-goals of more structured or "cookbook" style inquiries. Trouble-shooting coding problems over video call was actually found to be much easier when not having to crowd three or more people around a single laptop screen. Code snippets could be written on our own keyboards and shared over the video call's chat interface. Further, with the experiments built up in students' residences, the video calling saved them considerable time and problems incurred by breaking down, transporting, and again setting up equipment in the university building to demonstrate the setup to their peers and instructor. On the other hand, some types of instructor support were definitely less efficient via video call (e.g. helping to trouble-shoot a problem with electrical wiring or conveying ideas through a quick sketch). However, independent student researchers quickly learned the golden rule that any professional experimental scientist knows: if part of a setup is working, don't touch it!

Another factor contributing to resiliency that relates to AUC's curriculum structure is the offering of the Maker Lab as a stand-alone course over an extended four month timeframe. Since students' projects often required acquisition of new sensors or supplies, it was lucky that a single open-inquiry project was stretched out over 5.5 weeks to leave sufficient time for initial delivery of materials. Several projects also required taking measurements over many hours or days (which the Arduinos were set up to perform autonomously), meaning that iterative improvements on experimental methods required significant time. Additionally, it is hypothesized that students benefit from having time to sleep on or allow ideas to ferment, especially in the process of conducting empirical research.

Finally, the implementation of two full open-inquiry projects was advantageous for assessing individual students, especially given the decision to continue working mostly in pairs during the social distancing measures. The project pairs received grades on their assignments in a "first author – second author" format, whereby the first author was expected to lead and deliver the presentation and the second author played a supporting role whereby the assignment had a correspondingly smaller weight in their final grade. Because there were two projects, all students played the role of first author for all graded assignment types. In the circumstances of mandated social distancing, many project pairs naturally distributed the experimental work according to which team member would be the first author for the next presentation. But the first author for a particular presentation could count on their partner's support and critical feedback because their partner would share in the grade.

## Further considerations for pandemic-resiliency

While here in the Netherlands the pandemic control measures are now being refined and scaled back, there is still uncertainty about whether autumn semester courses can get back to normal, and if so, whether all students will be able to participate on campus. Planning a course like the Maker Lab for full pandemic resiliency requires some small adaptations in the first third of the course – which occurred this year before the start of the COVID-19 crisis. Further, one of us (CFJP) is working on adapting similar methods for a large first year physics lab course at the Delft University of Technology (TU Delft). The Maker Lab pilot and its abrupt transition to distance learning hold important lessons for optimizing the balance between independence, interaction, and personalized attention.

To enable online instruction of a Maker Lab-like course from day one, the tools may need to be delivered to (some) students at the beginning of the term, depending on whether collection on campus is possible.  Further, for rigid lockdown situations or in less-developed regions of the world, the initially distributed set of tools should include a larger array of sensors and supplies in case students are not able to quickly purchase sensors according to their project topics.  Of the 14 projects chosen by students this spring, half could have been served by a well-stocked initial set of tools, leaving lots of room for student choice and creativity.  Our colleagues at TU Delft have relevant experience as they run a hands-on design course in which ~150 students build prototypes, where each group requires different sensors, Arduinos, and other materials that must be distributed to them.

Another practical consideration is the elimination of the need for soldering via a combination of jumper wires/breadboards and a supply of common connectors/cables.  While the Maker Lab students were generally happy to have learned soldering skills, many did so the hard way via much trial and error, and these frustrations would have been exacerbated if they had not gotten in-class training in the first week of the course.

Further adaptations and improvements are now in development within a consortium of educators.  Besides adapting materials for different programs and levels, we are looking into possibilities for enabling more peer interactions and for including teaching assistants (TA's) in the teaching team.  The Maker Lab's original peer-feedback plans were substantially reduced after the transition to distance learning due to the perceived difficulty of holding regular synchronous class sessions.  However, this resulted in cutting sessions for everyone because a few could not (always) partake.  The Maker Lab course did not utilize TA's, so their potential roles and necessary training will require some consideration.  We expect that well-trained TA's will be able to individually conduct many of the individual team meetings, supervise peer-sharing and peer-feedback sessions, and fill in first drafts of student assessment forms.

Based on the achieved learning outcomes, and regardless of whether education is being offered on-campus or online, we recommend consideration of open-inquiry and flipped-classroom approaches, using modern technologies made accessible by the Maker movement.  We additionally hope our work can inform a "playbook" for transitioning physical science lab courses to fully online environments in case the COVID-19 pandemic continues, or in future circumstances that we now cannot foresee.  The Maker Lab and associated course materials have been published by Bradbury (2020). Interested readers are encouraged to contact the authors directly.

## Acknowledgments

We would like to acknowledge support in this project from the Comenius Fellowship grant from the Netherlands Initiative for Education Research (NRO) and the Dutch Ministry for Education (OCW) as well as AUC's Blending Learning funds and our supportive departments at AUC and TU Delft.  The Maker Lab pilot course greatly benefitted from input and help from Jasper Homminga, Gary Steele, Jan Koetsier, Thierry Slot, and especially co-instructor Paul Vlaanderen, and we are grateful for helpful discussions in the last month with Andy Buffler, Jean Heremans, and Thomas O'Donnell.